\documentstyle[prl,aps,preprint]{revtex}

\newcommand{\beq}{\begin{equation}}
\newcommand{\eeq}{\end{equation}}
\newcommand{\beqa}{\begin{eqnarray}}
\newcommand{\eeqa}{\end{eqnarray}}
\newcommand{\mpl}{M_{Pl}}
\newcommand{\lmk}{\left(}
\newcommand{\rmk}{\right)}
\newcommand{\lkk}{\left[}
\newcommand{\rkk}{\right]}

\newcommand{\rhov}{\rho_{\rm v}}
\newcommand{\ns}{|n\rangle}
\newcommand{\thetas}{|\theta\rangle}
\newcommand{\ps}{|+\rangle}
\newcommand{\ms}{|-\rangle}
\newcommand{\lp}{\langle +|}
\newcommand{\lm}{\langle -|}
\newcommand{\cH}{{\cal H}}

\begin{document}
%\begin{flushright} OU-TAP-166 \\
%\end{flushright}

\title{A cosmological constant from degenerate vacua}
\author{Jun'ichi Yokoyama}
\address{ Department
of Earth and Space Science, Graduate School of Science,\\ Osaka
University, Toyonaka 560-0043, Japan }
%\date{\today}
\maketitle
\tighten

\begin{abstract}
Under the hypothesis that the cosmological constant vanishes in the true
 ground state with lowest possible energy density, we argue that the
 observed small but finite vacuum-like energy density can be explained if
 we consider a theory with two or more degenerate perturbative vacua,
 which are unstable due to quantum tunneling, and if we still
 live in one of such states.  
An example is given making use of the topological vacua in
 non-Abelian gauge theories.

\end{abstract}

%\newpage
\vskip 1cm

Recent progress in observational cosmology has revealed that there are
{\it two} cosmological constant ($\Lambda$) problems.  One is the older
problem why the vacuum energy density is vanishingly small, or why
the intrinsic cosmological term cancels with accumulation of zero-point
energy in quantum field theory almost exactly.  
Observationally, the vacuum energy density 
$\rhov=3M_G^2\Lambda$
is no larger than the critical density $\rho_{cr0}=4\times 10^{-47}{\rm
GeV}^4=(3{\rm meV})^4$ today, where $M_G=\mpl/\sqrt{8\pi}=2.4\times 10^{18}$ GeV
is the reduced Planck scale.
On the other hand, 
since a natural cutoff scale of zero-point fluctuations of each quantum field 
is the Planck scale, we would expect $\left\langle {\rhov}
\right\rangle \simeq M_G^4$ from them, which is larger than the observational
constraint by a factor of $10^{120}$.
That is, the vanishingly small $\Lambda$ is realized as a result of a
cancellation of more than 120 digits \cite{review}.

The second, newer problem is that the above miraculous
cancellation does not seem to work in a perfect manner, that is,
there is increasing evidence that a finite positive component  still
remains in the vacuum-like energy density and that our Universe is in a
stage of accelerated expansion now.  For example, 
the analysis of SNIa data shows
that the probability that we live in a universe with $\Lambda=0$ is less
than one percent \cite{perl}.  It is undoubtedly one of the most important problems
in modern cosmology to explain the origin of this small but finite
density of vacuum energy.  For this purpose, we should keep in mind
the old problem, too.

Historically, a number of solutions have been proposed about the first
problem: adjustment mechanisms \cite{adj}, anthropic considerations
\cite{Anth,Anth2},
quantum cosmological approach \cite{HB,Col}, higher dimensional models
\cite{high}, 
and so on \cite{review,vil}. 
Among them, the quantum cosmological approach is based on the Euclidean
path integral of the wave function of the universe \cite{HH}.  
It has been claimed that such a path integral is dominated by the de
Sitter instanton solution proportional to $\exp(\frac{3\pi}{G\Lambda})$
and hence it is likely that the cosmological constant vanishes \cite{HB}.
Coleman further incorporated fluctuations of the spacetime topology in
terms of the ``wormhole'' configurations and found a double exponential
dependence \cite{Col}.  
One should note, however, that the expectation values obtained in these approaches
 should be regarded as giving an average over the time in
the history of the universe  \cite{review}.  So
they may not be necessarily related with the values we observe today.
We may rather interpret it as predicting a vanishing cosmological
constant in some ground state where the universe spends most of the time
in history.
\footnote{
We must also point out problems with Euclidean formulation of quantum
gravity, namely, positive-nondefiniteness of the Euclidean action, the
ambiguity of the signature of rotation \cite{rotation}, and the negativity of
the phase of the saddle point solution \cite{Polchinski}.}  
More recently, a number of higher dimensional models have been proposed
in which the maximally symmetric solution of the three-brane must have a
vanishing four-dimensional cosmological constant \cite{high}.  Since our Universe
has not settled in a maximally symmetric state, it is difficult to
understand implications of these results on current values of cosmological
parameters in our Universe, but we are tempted to 
interpret them as predicting a vanishing cosmological constant again in some
ultimate ground state.

In the present  Letter we argue the possible origin of the small
but finite cosmological constant without introducing any small numbers
under the hypothesis that 
the cosmological constant vanishes in the true
 ground state with lowest possible energy density.  In other words, we
attempt to solve the second problem under the assumption that the first
one is solved in the true ground state by arguing that we have not
fallen into that state.  Other proposed solutions to the second problem, 
such as quintessence \cite{q} (see also \cite{e} for earlier work) or
meV-scale false vacuum 
energy \cite{false}, are also based on such a hypothesis.

Our starting point is that the energy
eigenvalue of the true ground state of a theory with two or more degenerate
perturbative vacua, which cannot be transformed from one another without
costing energy,  is smaller than that of a quasi-ground state localized
around one of these states in field space by an exponentially small amount.
Here, by  perturbative vacua we mean a state with the lowest energy density
without taking possible tunneling effect to another perturbative
vacuum into account.  Our hypothesis is that the cosmological constant
vanishes {\it not} in these degenerate perturbative vacua but in the absolute ground
state with quantum tunneling effects taken into account, because there are both
classical and quantum contributions to the cosmological constant and
what we observe is their sum.

For illustration let us first consider an abstract field theory model 
whose perturbative vacuum states are classified into two distinct
categories labeled by $\ps$ and $\ms$ with $\langle +\ms = 0$ at the
lowest order.
We also assume that, although the transition from $\ps$ to $\ms$ is
classically forbidden, there is an instanton solution which describes
quantum tunneling from $\ps$ to $\ms$ and vice versa.  By nature the
instanton is localized
in space and (Euclidean) time with a finite Euclidean action
$S_0$.  Then the true ground state, $|S\rangle$, with this tunneling effect taken into
account, is given by the symmetric superposition of $\ps$ and $\ms$,
namely, 
\beq
|S\rangle = \frac{\ps+\ms}{\sqrt{2}}, \label{S}
\eeq
 where we have assumed that
$\ps$ and $\ms$ are normalized.

Now we evolve $|S\rangle$ with Euclidean time $T$ to
calculate $\langle S|e^{-\cH T}|S\rangle$ by summing up contributions of
instantons and anti-instantons as
\beqa
 \langle S|e^{-\cH T}|S\rangle &=& \frac{1}{2}\lmk\lp e^{-\cH T}\ps
 +\lm e^{-\cH T}\ms +\lp e^{-\cH T}\ms +\lm e^{-\cH T}\ps \rmk \nonumber \\
 &=& e^{-\rho_0VT}\sum_{j=0}^{\infty}\frac{1}{(2j)!}\lmk kVTe^{-S_0}\rmk^{2j}
+e^{-\rho_0VT}\sum_{j=0}^{\infty}\frac{1}{(2j+1)!}\lmk
kVTe^{-S_0}\rmk^{2j+1} \nonumber \\
 &=& \exp\lmk -\rho_0VT+kVTe^{-S_0}\rmk,  \label{Scalc}
\eeqa
where $\cH$ is the Hamiltonian, $k\equiv m^4$ is a positive constant and $V$
represents spatial volume \cite{asp}.  Here $\rho_0$ is the energy density of the
perturbative vacua, $\ps$ and $\ms$, which are presumably translationally
invariant.  Thus the energy density of the true ground state $|S\rangle$ 
is given by 
\beq
  \rho_S =\rho_0-m^4e^{-S_0}.  \label{Srho} 
\eeq
It is this energy density that vanishes under our hypothesis.  This in
turn implies that we find a nonvanishing vacuum energy density
\beq
\rho_0= m^4e^{-S_0} \label{0vac}
\eeq
in either of the perturbative vacuum states, $\ps$ or
$\ms$.  

Thus if our Universe is in one of the perturbative vacuum state
because it is too young to be relaxed into the true ground state
$|S\rangle$, we observe a nonvanishing vacuum energy density (\ref{0vac}) today.
Since the tunneling rate per unit volume
per unit time is given by $\Gamma \simeq m^4e^{-2S_0}$ apart from a
prefactor of order unity  \cite{decay}, the requirement that 
there should be no transition in the horizon volume in the cosmic
age reads
\beq
  \Gamma H_0^{-4} \simeq 9M_G^4/m^4 \lesssim 1,
\eeq
where $H_0^2 \cong \rho_0/(3M_G^2)$ is the current Hubble parameter
squared.  
We therefore find that, if
the parameters satisfy $m\gtrsim M_G$ and $S_0=
120\ln10+4\ln(m/M_G)$, we can account for the observed small value of the
cosmological constant without introducing any small numbers.

So far is a generic study to generate an exponentially small
difference in energy density using a theory with degenerate perturbative
vacua whose real ground state is given by their superposition.
Next in order to see how this mechanism may be implemented in 
 a more specific theory with this property,  let us
consider a famous example of an SU($N$) ($N\geq 2$) gauge theory 
whose perturbative vacuum states are
classified in terms of the winding number $n$ and denoted by
$\ns$ \cite{theta,jr}.  States with different winding numbers cannot be
transformed from each other by a continuous gauge transformation
\cite{jr} and
there is an energy barrier between them.  
Let us concentrate on the simplest case with $N=2$ hereafter.  An
instanton solution \cite{instanton}, which describes quantum tunneling
from one perturbative vacuum to another with
the change of the  winding number $\Delta n = 1$, 
can be expressed as
\beq
 A_\mu(x)=\frac{2R^2\eta_{a\mu\nu}(x_{\nu}-y_{\nu})\sigma^a}
{(x-y)^2\lkk (x-y)^2+R^2\rkk}, \label{instanton}
\eeq
where $\eta_{a\mu\nu}$ is the 't Hooft symbol \cite{thl}, $\sigma^a$ is the
Pauli matrix and $R$ is the size of
the instanton.  Here $y_\nu$ represents spacetime coordinates at its
center.  Thanks to the translational  and the scale invariance, the
Euclidean action does not depend on these quantities, namely,
\beq
S_0={1 \over {4g^2}}\int {d^4}xF_{\mu \nu }^aF^{a\mu \nu }={{8\pi^2 } \over {g^2}},
\eeq
where $F_{\mu \nu }^a=\partial _\mu A_\nu ^a-\partial _\nu A_\mu
^a+g\varepsilon ^{abc}A_\mu ^bA_\nu ^c$ is the field strength and
$g$ is the gauge coupling constant.

The true ground state in the presence of quantum tunneling is given
by an infinite sum of these $\ns$ states as
\beq
  \thetas =\sum_{n=-\infty}^{\infty}e^{in\theta}\ns,
\eeq
where $\theta$ is a real parameter \cite{theta,jr}.
One can easily find that this state is a real
eigenstate of the Hamiltonian $\cH$ in terms of the following
calculation
based on the dilute instanton approximation \cite{theta}.
\beqa
\left\langle {\theta '} \right|e^{-\cH T}\left| \theta  \right\rangle
&=&\sum\limits_{n,n'}  {\left\langle {n'} \right|e^{-\cH T}\left| n
\right\rangle }e^{in\theta -in'\theta '} \nonumber \\
  &=&\sum\limits_{n,n'} {\sum\limits_{m,\bar m} {{1 \over {m\! }}}{1
\over {\bar m\! }}\left( {KVTe^{-S_0}} \right)^{m+\bar m}\delta _{m-\bar
m,n-n'}}e^{in\theta -in'\theta '}  \label{thetaexp}  \\
  &=&\sum\limits_{n,n'} {e^{in(\theta -\theta ')}}\sum\limits_{m,\bar m}
{{1 \over {m\! }}}{1 \over {\bar m\! }}\left( {KVTe^{-S_0}}
\right)^{m+\bar m}e^{-i(m-\bar m)\theta } \nonumber \\
  &=& \exp \left( {2KVTe^{-S_0}\cos \theta } \right)
\delta \left( {\theta -\theta '} \right), \nonumber 
\eeqa
where $KVTe^{-S_0}$ represents  contribution of a single instanton or 
anti-instanton
and $m$ ($\bar m$) denotes  number of instanton (anti-instanton)
incorporated in each term, respectively.  Here $K$ is a positive constant and
$VT$ again represents spacetime volume.

Then the above equality (\ref{thetaexp}) clearly shows that each
$\theta$-vacuum $\thetas$ has a different energy density than the perturbative
vacuum $\ns$ by
\beq
\Delta\rho =-2Ke^{-S_0}\cos \theta . 
\eeq
Apparently the $\theta=0$ vacuum has the lowest energy (and no CP
violation), but one cannot immediately conclude that this is the only
vacuum state, because the other $\theta$-vacua are also stable against
gauge-invariant perturbations \cite{theta,jr}.  
Nonetheless under our hypothesis let us
take the CP-conserving $\theta=0$ vacuum with the lowest energy density
as the real ground state of the
full theory and assume that it is in this state that
the cosmological constant vanishes.  
For quantum gravitational approach to realize $\theta=0$ vacuum as the
ground state 
without introducing an axion, see \cite{Nielsen}.  With this
normalization of $\Lambda$, the vacuum energy density in each
perturbative vacua is found to be 
\beq
 \langle n| \rhov \ns = 2Ke^{-S_0}.
\eeq

In fact, the factor $K$, which is expressed as
\beq
  K=\frac{\pi^2}{4}\lmk\frac{8\pi}{g^2(\mu)}\rmk^4\int \frac{dR}{R^5}
 \exp\lmk -\frac{8\pi^2}{g^2(\mu)}+\frac{22}{3}\ln(\mu R)+6.998435\rmk, \label{Kk}
\eeq
in the case of the pure SU(2) gauge theory \cite{th}, 
 is  divergent due to the
contribution of arbitrary large instantons.  Here $\mu$ is a
renormalization scale.
In order to obtain a physical cutoff scale let us introduce
an SU(2) doublet scalar field $\Phi$ with a potential 
$V[\Phi]=\lambda(|\Phi|^2-M^2/2)^2/2$, following 't Hooft \cite{th}.
For $M \neq 0$, the solution (\ref{instanton}) is no longer an exact
one, but one could find an approximate solution, a constrained
instanton \cite{Affleck}, with the following properties. \\
(i) For $RM \lesssim 1$, the solution is given by (\ref{instanton}) and
\beq
  |\Phi(x)|=\lmk\frac{(x-y)^2}{(x-y)^2+R^2}\rmk^{1/2}\frac{M}{\sqrt{2}},
\eeq
for $(x-y)^2\lesssim M^{-2}$.  \\
(ii) For larger $(x-y)^2$, the solution rapidly approaches to the vacuum
values with 
\beq
  |\Phi(x)| - \frac{M}{\sqrt{2}} \sim e^{-\sqrt{\lambda}M|x-y|},~~~~~A_\mu(x) \sim
e^{-gM|x-y|} .
\eeq
(iii) The Euclidean action is finite and approximately given by
\beq
 S=\frac{8\pi^2}{g^2}+\pi^2R^2M^2.  \label{act}
\eeq

Using (\ref{act}) in (\ref{Kk}), the integral is now given by
\beq
  K=\frac{\pi^2}{4}\lmk\frac{8\pi}{g^2}\rmk^4\int \frac{dR}{R^5}
 \exp\lkk-\frac{8\pi^2}{g^2}-\pi^2R^2M^2+\frac{43}{6}\ln\lmk\frac{RM}{\sqrt{2}}\rmk
+6.759189\rkk,
\eeq
and we find,
\beqa
 \rhov &\cong& M^4\lmk\frac{8\pi}{g^2}\rmk^4 e^{-\frac{8\pi^2}{g^2}},
\label{rv} \\
 \Gamma &\simeq& M^4\lmk\frac{8\pi}{g^2}\rmk^4 e^{-\frac{16\pi^2}{g^2}}. \label{ma}
\eeqa
Demanding that $\rhov=10^{-120}M_G^4$ and that
the tunneling rate in the current horizon should be
smaller than unity in the cosmic age, $\Gamma H^{-4}_0 \lesssim 1$, we
find
\beqa
  \frac{\pi}{\alpha}+2\ln\alpha &=& 60\ln10 +2\ln\lmk\frac{M}{M_G}\rmk, \label{alpha}
\\
  M &\gtrsim & \alpha M_G,  \label{ineq}
\eeqa
where $\alpha\equiv g^2/(4\pi)$ is the coupling strength at the energy
scale  $M/\sqrt{2}$.
If the inequality (\ref{ineq}) is marginally satisfied, we find
$\alpha=1/44.4$ and $M=5\times 10^{16}$GeV.  If, on the other hand, we
take $M=M_G$ so that the cutoff scale of instanton is identical to the
presumed field-theory cutoff, we find $\alpha=1/47.$

Thus if our Universe happened to be created in a state with some
specific winding number and remains there up until now, we would observe
a nonvanishing vacuum energy density (\ref{rv}).  Although
$\theta$-vacuum is the real ground state of the theory, there is no {\it
a priori} reason that the Universe is created in this state.  
In fact,  in the possibly chaotic initial state of 
the early universe \cite{ci},
there may well be a small domain where the scalar field has a
nonvanishing expectation value with a constant SU(2) phase and
$A_{\mu}^a$ vanishes.  If such a region is exponentially stretched by
cosmological inflation \cite{inf} and our Universe is contained in
it, the state of our Universe would be more like the
perturbative vacuum $|n=0\rangle$ than some $\theta$-vacuum.  Then it is
not surprising that we observe a nonvanishing vacuum energy density
(\ref{rv}) today.  

Note that the Hubble parameter in the observable regime of
inflation, $H$, is constrained as
$H/(2\pi) \lesssim 4\times 10^{13}$GeV so that the tensor-induced anisotropy of
cosmic microwave background radiation \cite{gw} satisfies 
$\delta T/T \lesssim 10^{-5}$ \cite{COBE}.
Hence the amplitude of 
quantum fluctuations generated along the phase direction of the fields
during inflation is much smaller than
$M$, and it does not affect the realization of the state
$|n=0\rangle$.
Furthermore gravitational effects are negligibly small for the instanton
configuration even during inflation because $H \ll M$.
We also note that thermal transition to a state with a different winding
number is suppressed since the reheat temperature after inflation, $T_R$, is
typically much smaller than $M$ \cite{inf}. In fact, to avoid overproduction of
gravitinos, it should satisfy $T_R < 10^{12}$GeV \cite{moroi}.  
Hence we have only to worry
about quantum transition (\ref{ma}) as we have already done.

In summary, we have pointed out that in a field theory with two (or more)
degenerate perturbative vacua, the vacuum energy density of the true
ground state is smaller than that in a perturbative vacua by an
exponentially small amount if quantum tunneling between degenerate vacua
is possible, and that this may be utilized to explain the observed small 
value of the cosmological constant without introducing any small
quantities.

\acknowledgements{
The author is grateful to S.\ Mukohyama and S.\ Wada for useful communications.
This work was partially supported by the Monbukagakusho Grant-in-Aid for
Scientific Research No.\ 13640285  and ``Priority Area: Supersymmetry and
Unified Theory of Elementary Particles (No.\ 707).''}


\begin{thebibliography}{99}
\bibitem{review} S.\ Weinberg, Rev.\ Mod.\ Phys.\ {\bf 61}, 1 (1988).
\bibitem{perl}S.\ Perlimutter et al., Astrophys.\ J. {\bf 517}, 565
 (1999); B.P.\ Schmidt et al., Astrophys.\ J. {\bf 507}, 46 (1998);A.G.\
 Riess et al., Astron.\ J. {\bf 116}, 1009 (1998).
\bibitem{adj}A.D.\ Dolgov, in The very early universe, 
eds.\ G.W.\ Gibbons and S.T.\ Siklos (Cambridge, UK 1982).
\bibitem{Anth}
S.\ Weinberg, Phys.\ Rev.\ Lett. {\bf 59}, 2607 (1987);
A.\ Vilenkin, Phys.\ Rev.\ Lett. {\bf 74}, 846 (1995);
G.\ Efstathiou, Mon.\ Not.\ Roy.\ Astron.\ Soc. {\bf 274}, L73 (1995).
\bibitem{Anth2} For a review of anthropic considerations, see {\it e.g.}
J.D.\ Barrow and F.J.\ Tipler, Anthropic cosmological
 principle, (Oxford, UK 1988).
\bibitem{HB} E.~Baum, 
Phys.\ Lett.\ B {\bf 133}, 185 (1983); S.W.~Hawking,
Phys.\ Lett.\ B {\bf 134}, 403 (1984)
\bibitem{Col}S.R.~Coleman, 
Nucl.\ Phys.\ B {\bf 310}, 643 (1988).
\bibitem{high} S.\ Kachru, M.\ Schulz, and E.\ Silverstein, Phyr.\
 Rev. {\bf D62}, 045021 (2000); N.\ Arkani-Hamed, S.\ Dimopoulos, N.\
 Kaloper, and R.\ Sundrum, Phys.\ Lett.\ B {\bf 480}, 193 (2000);
S.-H.\ Henry Tye and I.\ Wasserman, Phys.\ Rev.\
 Lett. {\bf 86}, 1682 (2001).
\bibitem{vil} A.\ Vilenkin, hep-th/0106083.
\bibitem{HH}J.B.\ Hartle and S.W.\ Hawking, Phys.\ Rev.\ {\bf D28}, 2960 (1983).
\bibitem{rotation}A.D.\ Linde, Sov.\ Phys.\ JETP {\bf 60}, 211 (1984); 
A.\ Vilenkin, Phys.\ Rev.\  {\bf D30}, 509 (1984).
\bibitem{Polchinski}J.\ Polchinski, Phys.\ Lett.\ B {\bf 219}, 251
 (1989).
\bibitem{q}I.~Zlatev, L.~Wang and P.J.~Steinhardt,
Phys.\ Rev.\ Lett.\  {\bf 82}, 896 (1999); P.J.~Steinhardt, L.~Wang and
 I.~Zlatev, 
Phys.\ Rev.\ D {\bf 59}, 123504 (1999)
\bibitem{e}Y.\ Fujii, Phys.\ Rev.  {\bf D26}, 2580 (1982); L.H.\ Ford,
 Phys.\ Rev.  {\bf D35}, 2339 (1987); P.J.~Peebles and B.~Ratra,
Astrophys.\ J.\  {\bf 325}, L17 (1988);  Y.\ Fujii and T.\ Nishioka, Phys.\
 Rev.\ {\bf D42}, 361 (1990); K.\ Sato, N.\ Terasawa, and J.\ Yokoyama,
in Proc.\ XXIVth Recontre de Moriond, 
``The Quest for the Fundamental
Constants in Cosmology,'' eds.\ J.\ Audouze and J.\ Tran Thanh Van.
(Editions Fronti\`{e}res, France 1990) 193;  T.\ Nishioka and S.\ Wada, 
Int.\ J.\ Mod.\ Phys.  {\bf A8}, 3933 (1993).
\bibitem{false} N.\ Arkani-Hamed, L.J.\ Hall, C.\ Kolda, H.\ Murayama,
Phys.\ Rev.\ Lett. {\bf 85}, 4434 (2000);
S.M.\ Barr and  D.\ Seckel,  hep-ph/0106239;  
\bibitem{asp}S.~Coleman, Aspects of symmetry, (Cambridge, UK 1985).
\bibitem{decay}A.\ Lapedes and E.\ Mottola, Nucl.\ Phys.\ B {\bf 203},
 58 (1982). 
\bibitem{theta}C.G.\ Callan Jr, R.F.\ Dashen, and D.J.\ Gross, Phys.\
 Lett. {\bf 63B}, 334 (1976).
\bibitem{jr}R.\ Jackiw and C.\ Rebbi, Phys.\ Rev.\
 Lett. {\bf 37}, 172 (1976). 
\bibitem{instanton}A.M.\ Polyakov, Phys.\ Lett. {\bf 59B}, 82 (1975);
A.A.\ Belavin, A.M.\ Polyakov, A.S.\ Schvartz, and Yu.S.\ Tyupkin,
 Phys.\ Lett. {\bf 59B}, 85 (1975).
\bibitem{thl}G.\ 't Hooft, Phys.\ Rev.\ Lett. {\bf 37}, 8 (1976). 
\bibitem{Nielsen}H.B.\ Nielsen and M.\ Ninomiya, Phys.\ Rev.\ Lett. {\bf
 62}, 1429 (1989).
\bibitem{th}G.\  't Hooft, Phys.\ Rev. {\bf D14}, 3432 (1976); {\bf
 D18}, 2199 (1978) (E).
\bibitem{Affleck}I.\ Affleck, Nucl.\ Phys.\ B {\bf 191}, 429
 (1981); M.\ Nielsen and N.K.\ Nielsen, Phys.\ Rev. {\bf D61}, 105020
 (2000).
\bibitem{ci}
 A.D.\ Linde,  Phys.\ Lett. \textbf{129B}, 177(1983).
\bibitem{inf}
For a review of inflation, see, e.g.\ A.D.\ Linde, Particle Physics and
Inflationary
Cosmology (Harwood, Chur, Switzerland, 1990); K.A.\ Olive, Phys.\ Rep. {\bf
190},181(1990); A.R.\ Liddle and D.H.\ Lyth, Cosmological inflation and
 large-scale structure (Cambridge, UK, 2000).
\bibitem{gw}V.A.\ Rubakov, M.V.\ Sazin, and A.V.\ Veryaskin, Phys.\
 Lett. {\bf 115B}, 189(1982); L.F.\ Abbott and M.B.\ Wise, Nucl.\ Phys.\
 B {\bf 244}, 541 (1984).
\bibitem{COBE}
C.L.\ Bennet {\it etal}., Astrophys.\ J.\ Lett. {\bf 464}, L1(1996).
\bibitem{moroi}M.\ Kawasaki and T.\ Moroi, Prog\ Theor.\ Phys. {\bf 93},
 879 (1995).

\end{thebibliography}
\end{document}